\documentclass[aps,prl,superscriptaddress,floatfix,reprint,longbibliography]{revtex4-2}
\usepackage{amsmath}
\usepackage{tikz}
\usetikzlibrary{arrows}
\usetikzlibrary{decorations.pathmorphing}
\usepackage{graphicx} 
\usepackage{amssymb}
\usepackage{hyperref}
\usepackage{url}
\usepackage{subfigure}
\usepackage{bbm}
\usepackage{hyperref} 
\usepackage{graphicx}% Include figure files
\usepackage{dcolumn}% Align table columns on decimal point
\usepackage{bm}% bold math
\usepackage{amsmath}
\usepackage[utf8]{inputenc}
\usepackage{CJKutf8}

\newcounter{amg}

\begin{document}

%\title{Exceptional points at the PT transition in SYK Liouvillians}

\title{Role of exceptional points in the dynamics of the Lindblad Sachdev-Ye-Kitaev model}

%%% Alternative title

%\title{Exceptional points and anomalous equilibration at the weak to strong coupling transition in SYK LIouvillians}
\author{\begin{CJK*}{UTF8}{gbsn}
		Jie-ping Zheng (郑杰平)
\end{CJK*}}
\email{jpzheng@sjtu.edu.cn}
\affiliation{Shanghai Center for Complex Physics, School of Physics and Astronomy, Shanghai Jiao Tong University, Shanghai 200240, China}
\author{Jorge Dukelsky}%
\email{j.dukelsky@csic.es}
\affiliation{Instituto de Estructura de la Materia, IEM-CSIC, Serrano, 123 E-28006 Madrid, Spain}
\author{Rafael A. Molina}
\email{rafael.molina@csic.es}
\affiliation{Instituto de Estructura de la Materia, IEM-CSIC, Serrano, 123 E-28006 Madrid, Spain}
% \author{Jie Ping Zheng}
\author{Antonio M. Garc\'ia-Garc\'ia}
\email{amgg@sjtu.edu.cn}
\affiliation{Shanghai Center for Complex Physics, School of Physics and Astronomy,
	Shanghai Jiao Tong University, Shanghai 200240, China}

%\affiliation{
%\affiliation{%


\date{\today}% It is always \today, today,

\begin{abstract}

The out of equilibrium dynamics of the Sachdev-Ye-Kitaev model (SYK), comprising $N$ Majoranas with random all-to-all four-body interactions, minimally coupled to a Markovian bath modeled by the Lindblad formalism, displays intriguing nontrivial features. In particular, the decay rate towards the steady state is a non-monotonic function of the bath coupling $\mu$, and an analogue of the Loschmidt echo for dissipative
quantum systems undergoes a first order dynamical phase transitions that eventually becomes a crossover for sufficiently  large $\mu$. We provide evidence that these features have their origin in the presence of exceptional points in the purely real eigenvalues of the SYK Liouvillian closest to the zero eigenvalue associated with the steady state. An analytic calculation at small $N$, supported by numerical results for larger $N$, reveals that the value of $\mu \sim 0.1$ at which the exceptional point corresponding to the longest living modes occurs is close to a local maximum of the decay rate. This value marks the start of a region of anomalous equilibration where the relaxation rate diminishes as the coupling to the bath becomes stronger. Moreover, the mentioned change from transition to crossover in the Loschmidt echo occurs at a larger $\mu \sim 0.3$ corresponding with a proliferation of exceptional points in the low energy limit of the Liouvillian spectrum. We expect these features to be generic in the approach to equilibrium in quantum strongly interacting many-body Liouvillians.

\end{abstract}

\maketitle
 
 %In recent years, the quantum thermalization of isolated many-body systems has emerged as a central topic in non-equilibrium statistical mechanics, driven both by fundamental interest and experimental progress in ultracold atomic systems and other controllable quantum mechanical platforms \cite{Greiner2002,Bloch2008,Linden2009,Gogolin2016,Neill2016,Binder2018,Campbell2025}. Despite the unitary nature of quantum dynamics, generic closed systems often appear to relax towards thermal equilibrium, where local observables are well-described by thermal ensembles. A key ingredient underlying this emergent thermal behavior is quantum chaos: in chaotic systems, the eigenstate thermalization hypothesis (ETH) provides a microscopic foundation for thermalization by postulating that individual energy eigenstates already encode thermal properties \cite{Deutsch1991,Srednicki1994,Rigol2008,Polkovnikov2011}. The onset of chaos governs not only whether a system thermalizes, but how rapidly it does so. Specifically, the decay rate of local observables toward their thermal values - and more generically, the system's approach to equilibrium - is closely tied to spectral properties and dynamical signatures of chaos, such as level statistics and operator spreading \cite{DAlessio2016,Bertini2015,Swingle2018}. Understanding the interplay between chaos, entanglement growth, and thermalization remains an active area of research, with implications for quantum simulation, information scrambling, and the limits of ergodicity in many-body systems \cite{Gogolin2011,Turner2018,Seetharam2018,Abanin2019}.

A central problem in modern theoretical physics is to reconcile the unitary evolution of many-body quantum chaotic systems with the irreversible nature of the thermal equilibrium state resulting from this temporal evolution. 
%In classical systems this problem can be traced back to the foundation of statistical mechanics by Boltzmann \cite{Tolman1938}. The most accepted solution is that thermalization arises naturally from ergodic dynamics and coarse-graining \cite{Gallavotti1999}. On the other hand, quantum systems evolve under deterministic, reversible dynamics that conserve the purity of the wavefunction. However, at the same time, local observables in non-integrable closed systems often exhibit relaxation towards thermal values, suggesting the emergence of statistical behavior from purely coherent dynamics. 
%This raises a fundamental question: how can thermodynamic irreversibility arise from microscopic unitarity? 
One promising strategy to address this puzzle involves studying quantum systems weakly coupled to an external environment, in the limit where the coupling strength approaches to zero. This regime allows the observation of the interplay between intrinsic thermalization mechanisms - such as those rooted in the unitary quantum chaotic dynamics, or more generally in the applicability of the eigenstate thermalization hypothesis \cite{srednicki2002},  and the onset of dissipative effects. A particularly insightful approach in this direction is the study of the Loschmidt echo, which quantifies the sensitivity of the dynamics to small perturbations and has been shown to exhibit a perturbation independent regime both in Nuclear Magnetic Resonance (NMR) experiments \cite{Levstein1998} and in semiclassical chaotic systems \cite{Jalabert2001,Goussev2012}. Crucially, the weak-coupling limit can act as a controlled probe: it permits the system to evolve predominantly under its unitary dynamics, while the environment gradually reveals information about its approach to equilibrium. By tuning the coupling to be sufficiently small, one may extract signatures of internal equilibrium that are otherwise inaccessible, effectively using the environment as a diagnostic tool rather than a driver of thermalization. 

%This perspective opens a potential path toward resolving the apparent contradiction between unitary evolution and thermal behavior, by framing open-system dynamics not as a competing mechanism but as a window into the intrinsic relaxation processes of isolated quantum matter. 

%A crucial caveat is that the weak-coupling limit and the thermodynamic limit generally do not commute in Lindblad dynamics \cite{Znidaric2015}. 
%This non-commutativity can signal a breakdown of the bath–system description, since the Born–Markov approximation underlying the Lindblad equation assumes that bath correlation times are much shorter than the relevant system timescales. 
%In the thermodynamic limit, however, the system level spacing becomes vanishingly small, and this separation of timescales may no longer hold \cite{Tupkary2022}. 
A problem with this approach is that the limits of weak coupling and the thermodynamic limit do not commute \cite{Mori2024}.  Fortunately, a careful finite-size scaling analysis makes possible the extrapolation of results that enables us to extract intrinsic relaxation properties of the many-body system itself \cite{Mori2024} despite being coupled to a bath.

Building on this perspective, we aim to explore the full crossover between the weak and strong coupling to the bath regimes in quantum chaotic systems, with the aim of understanding how intrinsic thermalization mechanisms interact with, and eventually give way to, environment-induced equilibration. 

%While the weak-coupling limit allows for the observation of unitary relaxation processes before environmental effects dominate, increasing the system-bath coupling strength introduces a gradual shift in the origin and nature of thermalization. In the strong-coupling regime, the dynamics is no longer governed primarily by the system’s internal Hamiltonian, but instead by the structure of the environment and the form of the system-environment interaction. 
%This problem raises a host of challenging questions: How do signatures of quantum chaos — such as level statistics, entanglement growth, or operator spreading — evolve as coupling increases? At what point does the system cease to exhibit intrinsic thermalization and instead thermalize purely due to dissipation? Are there universal dynamical signatures characterizing this crossover, or do they depend sensitively on the spectral and coupling properties of both the system and the bath?  By systematically studying this transition across coupling strengths, 
%More specifically, we aim to 
%build a unified understanding of quantum thermalization that bridges the gap between isolated and open systems, 
%shedding light on how precisely chaos, coherence, and dissipation conspire to achieve thermalization. 
A main tool in our analysis will be the study of exceptional points (EP) \cite{kato1965} in the spectrum of the Liouvillian that governs the dynamics of quantum systems coupled to a bath.  
EP are non-Hermitian degeneracies where both eigenvalues and eigenvectors coalesce.
% giving rise to novel physical phenomena absent in Hermitian systems. 
Around these EP the spectrum becomes very sensitive to small changes to the tuning parameter which has a profound effect
on a broad range of observables \cite{Heiss2004,Heiss2012,Ashida2020}.
Exceptional points were first confirmed experimentally in optics \cite{dembowski2001,guo2010} but later they have been studied in a variety of contexts including cold atoms \cite{li2019}, superconductivity \cite{kawabata2018}, entanglement and quantum information \cite{lee2014,kawabata2018}, open Markovian systems \cite{hatano2019,Khandelwal2021,Rubio2022b}, topological quantum matter \cite{Heiss2012,Molina2018,Miri2019} and in applications ranging from enhancing detector sensitivity \cite{wiersig2014,Hodaei2017,Rosa2021,Wong2023} to unconventional lasing \cite{ElGanainy2018,Ozdemir2019}. 
 
For the purpose of this study, we shall focus on the Sachdev-Ye-Kitaev model (SYK), $N$ Majorana fermions in zero spatial dimensions with all-to-all interactions,   \cite{french1970,french1971,bohigas1971,bohigas1971a,benet2001,sachdev1993,kitaev2015,maldacena2016} coupled to a bath described by the Lindblad formalism \cite{lindblad1976,gorini1976}. The out of equilibrium dynamics of this model have already been studied \cite{sa2022,kulkarni2022,kawabata2023,GarciaGarcia2023} in the large $N$ limit by using path integral techniques. A non-monotonic dependence of the equilibration time with the coupling to the bath \cite{GarciaGarcia2023} and a rich pattern of dynamical phase transitions \cite{kawabata2023,GarciaGarcia2023} are observed. 

{\em Model:} 
The Liouvillian equation for the evolution of the density matrix $\rho$ for an open quantum system is
\begin{equation}\label{eq:Liouville}
    \frac{d\rho}{dt}=\cal{L}(\rho).
\end{equation}
We are interested in the dynamics of the Sachdev-Ye-Kitaev model \cite{kitaev2015,maldacena2016,sachdev1993} coupled to a Markovian bath, termed Lindblad SYK \cite{sa2022,kulkarni2022,GarciaGarcia2023}, whose Liouvillian using the Lindblad formalism is given by:
\begin{equation}\label{eq:SYKLindblad}
{\cal{L}}(\rho) = -i\left[H_{SYK},\rho\right]+\sum_{\alpha}L_{\alpha}\rho  L_{\alpha}^{\dagger}-\frac{1}{2}\left\lbrace L_{\alpha}^{\dagger}L_{\alpha},\rho \right\rbrace.   
\end{equation}
where $H_{SYK}$ stands for the SYK Hamiltonian with $q$-body interactions and $N$ Majoranas operators $\psi_i$, 

%The Hamiltonian part is given by the SYK Hamiltonian with $q=4$-body interactions.

%\begin{equation}\label{eq:Hamiltonian}
%    H_{SYK}=-i\sum_{i_{1}<...<i_{4}} J_{i_1 i_2 i_3 i_4} \psi_{i_1}\psi_{i_2}\psi_{i_3}\psi_{i_4},
%\end{equation}

\begin{equation}\label{eq:Hamiltonian}
    H_{SYK}=i^{q/2}\sum_{i_{1}<...<i_{q}} J_{i_1 ... i_q} \psi_{i_1}...\psi_{i_q},
\end{equation}
with  $\left\lbrace\psi_i,\psi_j\right\rbrace=\delta_{ij}$.
The different couplings $J_{i_1 ... i_q}$ are Gaussian distributed random numbers with zero mean and variance $\sigma^2=(q-1)!/N^{q-1}$.  We will focus on the case $q=4$ whose dynamics is quantum chaotic and shows the anomalous equilibration \cite{GarciaGarcia2023} we are interested in. 
For the dissipative part we consider linear jump operators of the form   $L^i=\sqrt{\mu}\psi_i.$
%\begin{equation}\label{eq:jumps}
%    L^i=\sqrt{\mu}\psi_i.
%\end{equation}
Instead of treating the Liouvillian equation in its original form with the density matrix, we use a vectorization procedure where we made a mapping of the density matrices to a vector living in a doubled Hilbert space composed of the tensor product of the original Hilbert space and its dual $\cal{H}^+ \otimes \cal{H}^-$ \cite{kulkarni2022},
\begin{eqnarray}\label{eq:lio}
    {\cal{L}}&= -iH^+_{SYK} \otimes {\cal{I}}^-+i(-1)^{q/2}{\cal{I}}^+ \otimes H^-_{SYK} \nonumber \\ & +i\mu\sum_i \psi_i^+ \psi_i^- -\mu \frac{N}{2} {\cal{I}}^+ \otimes {\cal{I}}^-.
\end{eqnarray}
The eigenvalues $\lambda$ of Liouvillians are real or complex conjugate due to the Hermiticity of the density matrix \cite{minganti2018}. 
%In our case, 
%This can be shown explicitly by noticing the PT-symmetry of the vectorized Liouvillian. 
For Liouvillians in Lindblad form there is always, at least, one steady state of zero eigenvalue. For our model, it is easy to prove \cite{GarciaGarcia2023} that the steady state corresponds to the infinite temperature density matrix. 

It is also possible to demonstrate that the vectorized Liovillian Eq.~(\ref{eq:lio}) has parity symmetry that allows to write it  in a block diagonal form. One of the blocks has the diagonal components of the density matrix. The infinite-temperature steady state with zero eigenvalue is obtained from the diagonalization of this block. The other block has only non-diagonal components of the density matrix. A quantity of great importance in the dynamics of open quantum systems is the dissipative gap which is defined as the real
part of the eigenvalue closest to the steady state. Therefore, it is also the slowest decay rate $\Gamma_0$ in the dynamics. Its inverse is the typical equilibration time after a weak perturbation. The state in this latter block with the smallest absolute value of the real part defines the dissipative gap controlling the decay of coherences. In the following, we show for small $N$ that this state undergoes a real-complex transition going through an EP as a function of the coupling strength to the bath $\mu$.

 %%%%%% Comments on parity symmetry and its meaning in general and for the dissipative gap

%\section{Analytical calculation of exceptional points for $N = 4$}
%\label{sec:Minex}

\begin{figure}[htbp]
    \centering 
    \includegraphics[width=0.49\columnwidth]{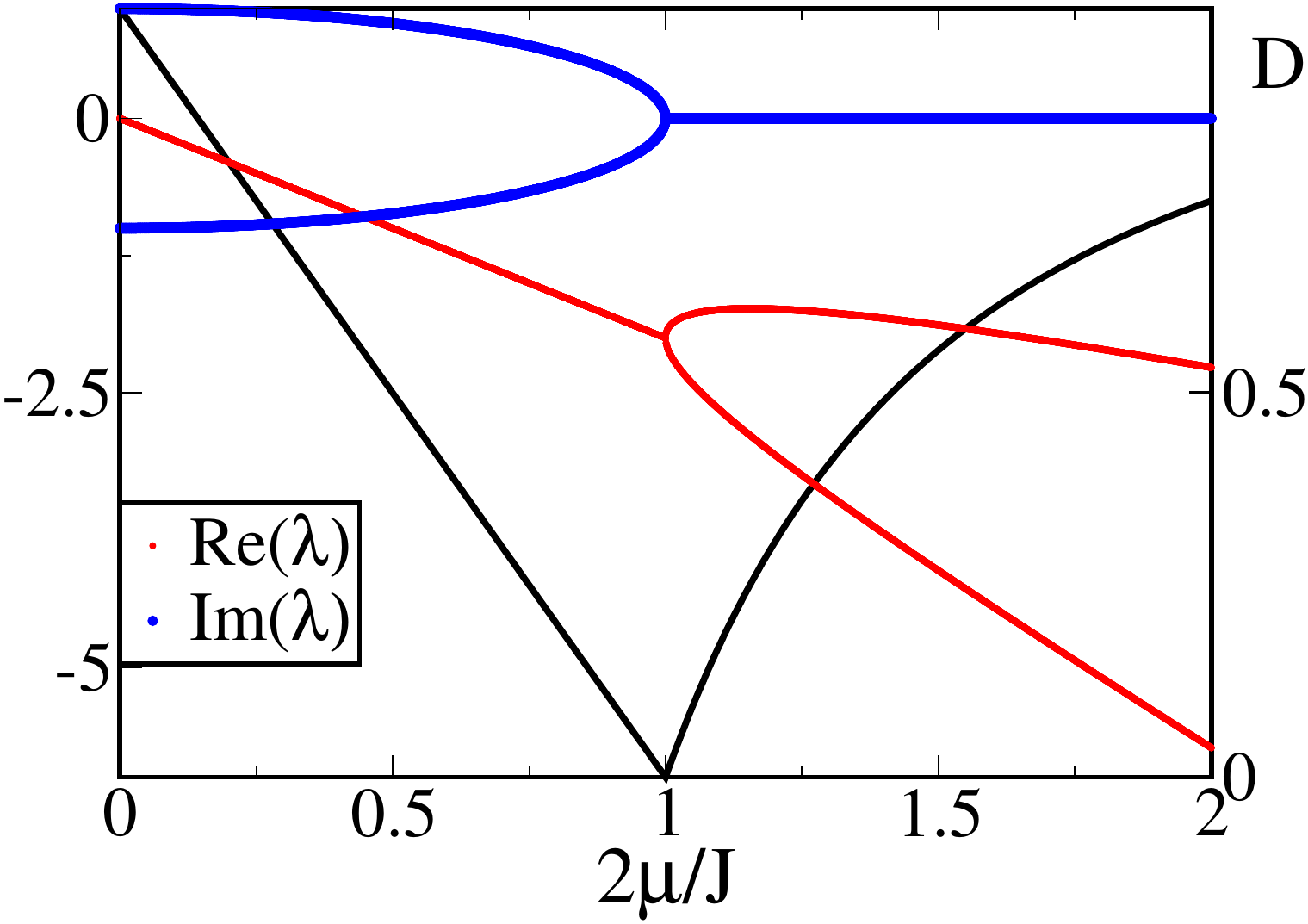}
    \includegraphics[width=0.49\columnwidth]{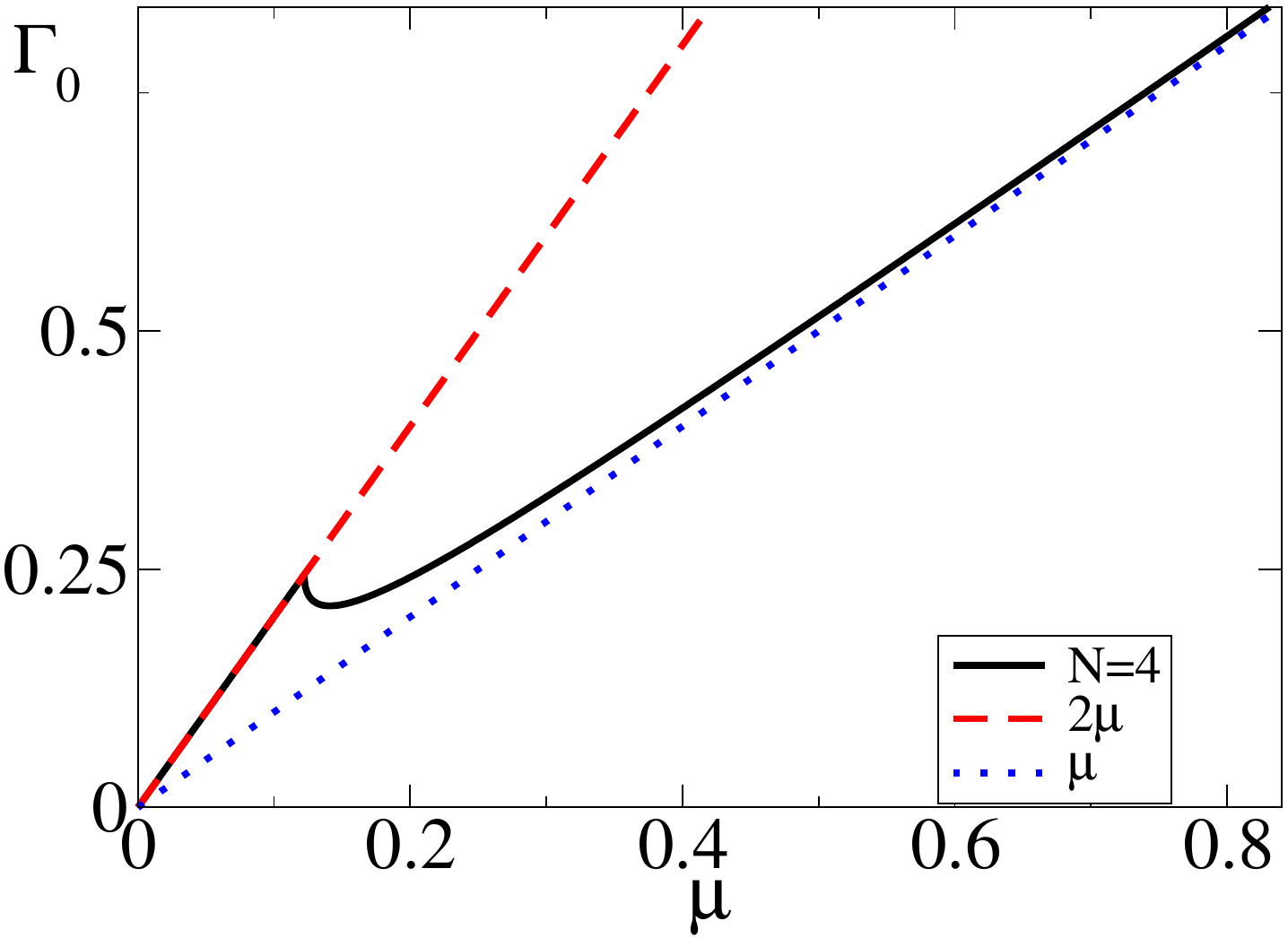}
    \caption{ {\em Left}: Blue and red lines stand for real and imaginary components of the complex eigenvalues as a function of $2\mu/J$ in the case of $q = N = 4$ Majorana fermions. The black line stands for the distance between the two eigenstates defined as $D=1-\left|\left<\Psi_1|\Psi_2\right>\right|$, where $\left<\Psi_1|\Psi_2\right>$ is defined as the scalar product of the two right eigenvectors, as a function of $2\mu/J$ for the same case, showing the coalescence for $\mu = J/2$. {\em Right}: $\Gamma_0$ Eq.~(\ref{eq:gamma0}) as a function of the coupling to the bath for $q = 4$. % Left: Analytic decay rate $\Gamma_0$ Eq.(\ref{eq:gamma0}). 
We have used a value of $J \approx 0.244$, corresponding to the average value of $|J|$ for $N=4$, according to the scaling used for the distribution of the constants $J$ in the $H_{SYK}$ Hamiltonian. }\label{fig:minmodel}
\end{figure}

{\em Analytical calculation of exceptional points for $N=4$:} We initiate our analysis with the exact calculation of the Liouvillian spectrum for $N = 4$ in order to provide evidence of the existence of EPs. 
%We start analytically solving a minimal example that already shows the presence and the %effect of exceptional points as a function of the coupling constant $\mu$.
%When the number of Majorana fermions is equal to 4, the Hamiltonian $H_{SYK}$ is diagonal %(in the chosen basis). The Liouvillian, however, is not. In this minimal case, the %Hamiltonian has a dimensionality of 4, so the Liouvillian in vectorized form has a %dimensionality of 16. As we mentioned earlier, the number of fermions is not conserved but %is conserved the parity of the number of fermions. %The odd case is trivial and the only %non-trivial case is the even case with a matrix dimension of 8. 
The use of the appropriate basis exploiting the parity symmetry of the model allows to write down the $N = 4$ vectorized Liouvillian Eq.~(\ref{eq:lio}), a $2^N \times  2^N$ matrix, in different blocks. The relevant ones for the dissipative gap are a trivial block with a zero eigenvalue and a $2 \times 2$ block,
\begin{equation}\label{eq:4Maj}
{\cal L}_{gap}=\begin{pmatrix}
-\frac{J}{2}i -2\mu & -\mu \\
-\mu & \frac{J}{2}i -2\mu
\end{pmatrix}
\end{equation}
where $J=J_{1234}$ is the random coupling in Eq. (\ref{eq:Hamiltonian}), which for $N=4$ is the only coupling since $N=q$. The two different eigenvalues of this matrix are $\lambda=-2\mu\pm \sqrt{\mu^2-(\frac{J}{2})^2}$ which implies a transition from complex conjugates eigenvalues for $\mu<J/2$ to real eigenvalues for $\mu>J/2$, see Fig. \ref{fig:minmodel}. The degeneracy at $\mu=J/2$ is the so called EP where the eigenvalue of the Liouvillian is degenerate but, unlike the Hermitian counterpart, the eigenvectors coalesce, namely, they merge into a single eigenvector and the Liouvillian becomes non-diagonlizable.

%Note that for non-Hermitian operators the right and left eigenvectors are different and the %right eigenvectors are not orthogonal, in general.
%In particular, the eigenvector corresponding to $\lambda=-2\mu+\sqrt{4\mu^2-J^2}$ is  while the eigenvector corresponding to $\lambda=-2\mu-\sqrt{4\mu^2-J^2}$ is $$, the distance between them defined as approaches as $\mu=$, the position of the exceptional point. Note that for non-Hermitian operators, ...  
In Fig. \ref{fig:minmodel} (left plot, black line), we depict explicitly this eigenvector coalescence by showing that the distance $D=1-\left|\left<\Psi_1|\Psi_2\right>\right|$, where $\langle \ldots \rangle$ stands for scalar product, between the right eigenvectors corresponding to the two coalescing eigenvalues going to zero as the coupling to the bath $\mu$ approaches the EP. 

Having demonstrated the presence of exceptional points in our model, we now proceed to investigate their role in the out of equilibrium dynamics. For $N=4$, we can compute the dissipative gap $\Gamma_0$ analytically as it is just minus the real eigenvalue in the upper branch in the left panel of Fig.~(\ref{fig:minmodel}),  
%\textcolor{red}{No se bien a que figura/curva te refieres. Hay que cambiar top por left or right}.
\begin{equation}
    \Gamma_0=\left\{
    \begin{array}{cc}
           2\mu & \,\,\, \mu < J/2 \\
           2\mu-\sqrt{\mu^2-(\frac{J}{2})^2} & \,\,\, \mu>J/2
    \end{array} \right\}
    \label{eq:gamma0}
\end{equation}
which is shown (black line) in the right panel of Fig. \ref{fig:minmodel}. We observe a linear increase $\Gamma_0=2\mu$ for small values of $\mu$ up to the value of $\mu$ corresponding to the EP (dashed red line in Fig.~\ref{fig:minmodel}). Then, there is a region of anomalous behavior of $\Gamma_0$ in which increasing the coupling to the environment decreases $\Gamma_0$, and, thus, increases the equilibration time. Asymptotically, we retrieve a linear increase of $\Gamma_0$ but with a different slope, $\Gamma_0 \approx \mu$ (blue dotted line in Fig.~\ref{fig:minmodel}). 

{\em Size scaling to the thermodynamic limit of the dissipative gap}:
The analytic results presented in the previous section and its interpretation, which is the core of this work, are now further supported by the analysis of the spectrum of the Liouvillian Eq.~(\ref{eq:lio}) obtained by exact diagonalization techniques for larger values of the number of Majoranas $N$. We compute again the decay rate $\Gamma_0$ as the dissipative gap in the spectrum by computing the eigenvalue with the real part closest to zero of the block with opposite parity to the steady state block. We use the Lanczos-Arnoldy method for non-Hermitian systems to be able to go up to system sizes involving $N_{\rm tot}=44$ Majoranas, namely, $N=22$ Majoranas in each copy of the vectorized Liouvillian. We then perform ensemble average and extrapolate our results to the thermodynamic limit by doing a simple linear finite size scaling $\Gamma_0(N_{\rm tot}) = \Gamma_0(\mu) + b/N_{\rm tot}$ with $b, \Gamma_0(\mu)$ as fitting parameters. For that purpose, we employ results for $N_{\rm tot} = 24, 28, 32, 36, 40, 44$ and $\mu \in [0.05,0.4]$. For $N_{\rm tot} = 44$, we could only perform ensemble average over about five disorder realizations. The error in our estimation of $\Gamma_0(\mu)$ comes from both the
fitting procedure for the size scaling and the variance of ensemble average. We have observed that for the smallest $\mu = 0.05$, the value of $\Gamma_0$ is sensitive to considering, or not, the lowest $N_{\rm tot} = 24, 28$. For the sake of consistency we have still employed these two values in the fittings. Note that we cannot explore smaller values of the coupling $\mu$ because the mentioned non-commutativity of the $N \to \infty$ and $\mu \to 0$ limits  which restricts \cite{Mori2024} $\mu \gtrsim 1/N$. 
 
\begin{figure}[htb]
	\centering
	\includegraphics[width=0.49\columnwidth]{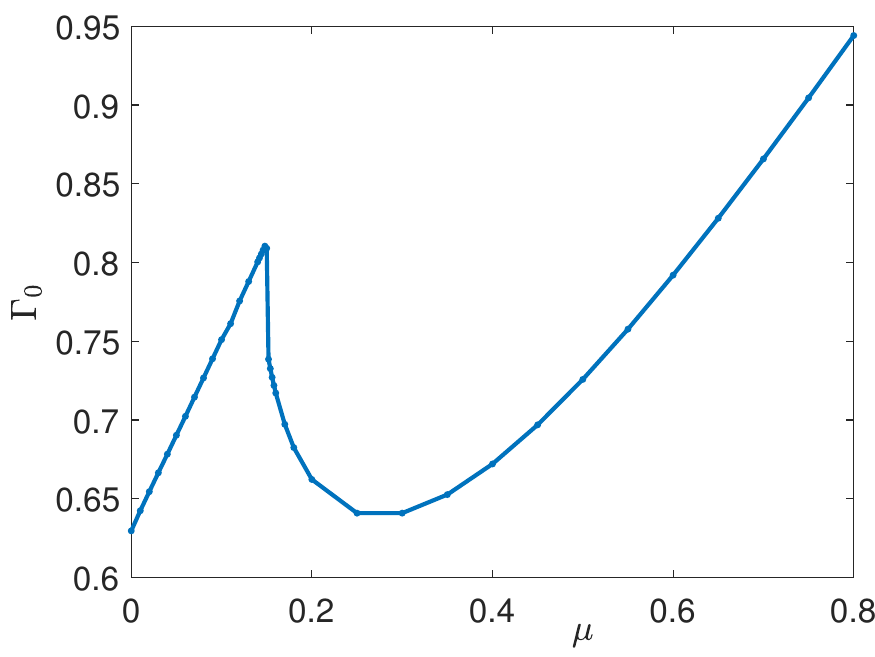}
        \includegraphics[width=0.49\columnwidth]{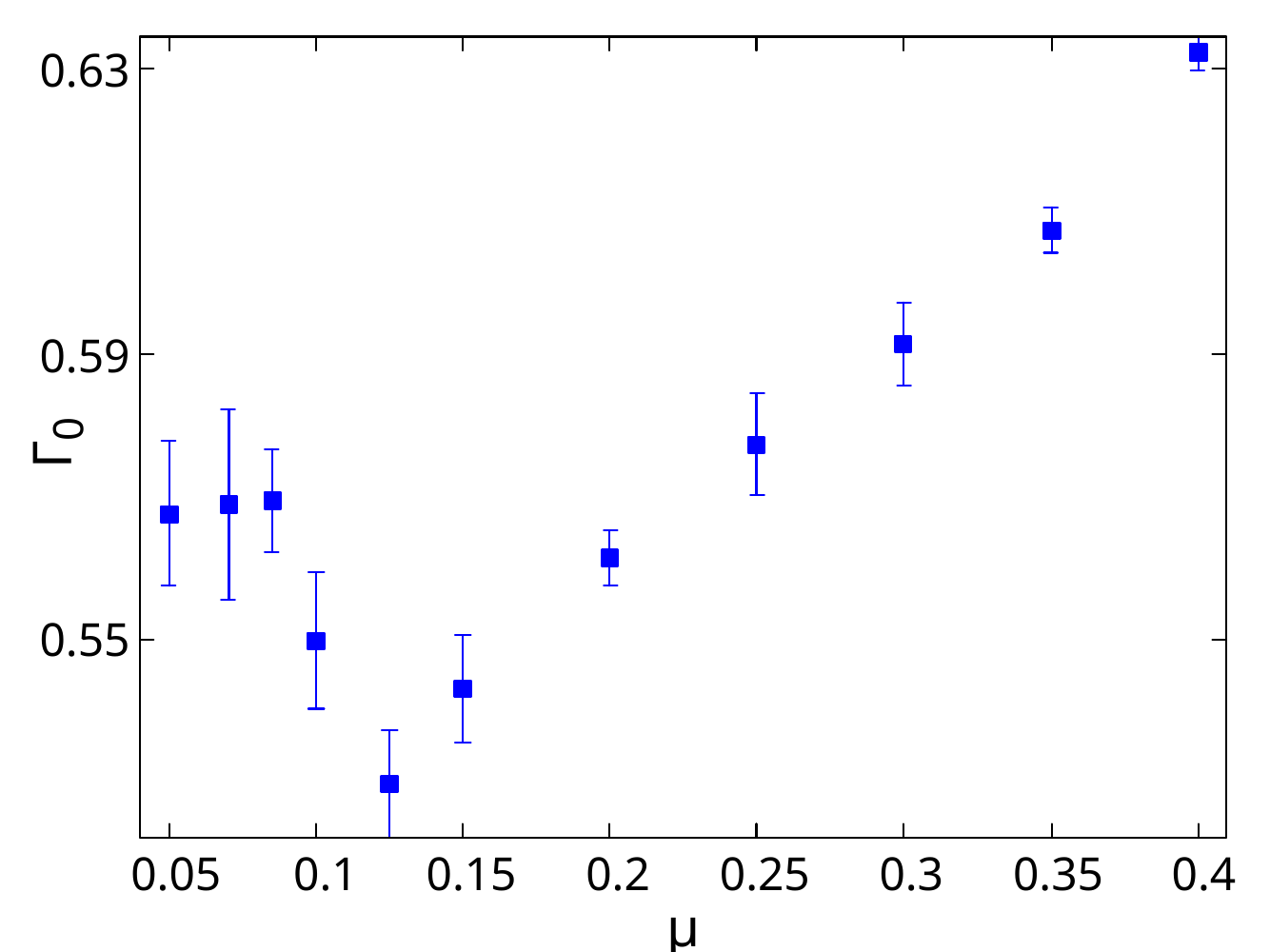}
	\caption{Comparison between the decay rate $\Gamma_0$ computed (left) from the Green's function in the large $N$ limit, see End Matter for details, and the numerical calculation of the spectral gap (right) employing exact diagonalization and performing a finite size scaling analysis.} 
	\label{fig:gap_tl}
\end{figure}

Results depicted in Fig.~\ref{fig:gap_tl} show the comparison of $\Gamma_0$ employing this procedure (right panel) with $\Gamma_0$ obtained previously by some of us \cite{GarciaGarcia2023} (left panel) by fitting the exponential decay rate $\Gamma_0$ of the retarded Green's function in the large $N$ limit, see End Matter for a additional details of this calculation. 
 By definition, this decay rate $\Gamma_0$ is the inverse of the typical equilibration time towards the steady state at infinity temperature.
 %The precise definition of the Green's function together with details of its calculation in the mentioned large $N$ limit are found in the End Matter. % \ref{sec:endmatter}.   
%The Keldysh path integral offers a systematic way to treat the nonequilibrium dynamics of many-body systems under the Lindblad equation. By formulating the evolution of the density matrix along forward and backward time branches, it doubles the degrees of freedom and maps the problem onto a state in a doubled Hilbert space much like the vectorization process we have used to solve our Lindblad equation. This representation makes it possible to apply field-theoretical methods to capture both coherent and dissipative processes, from transient dynamics to steady states. The results in the large N theory come from a calculation of a fitting of the exponential decay of the Green function with respect to time in the long time limit. 

\begin{figure}
	\centering
	%\subfigure[]{\includegraphics[width=.45\columnwidth]{iS_q_4_ld_05_t_1_20}}
	%\subfigure[]{\includegraphics[width=.45\columnwidth]{iS_q_4_ld_p1_t_1_20}}
~	\subfigure[]{\includegraphics[width=.49\columnwidth]{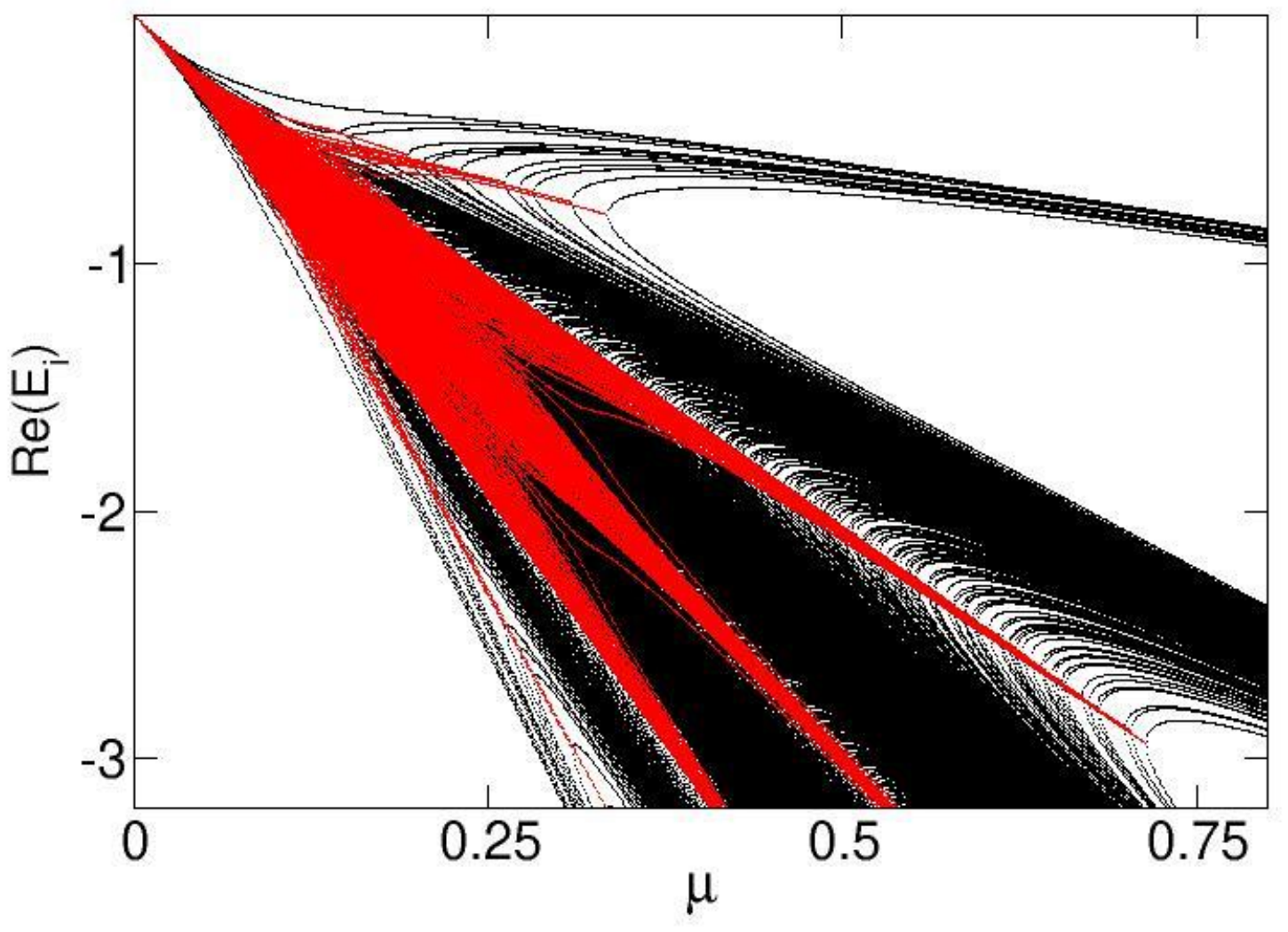}}
	\subfigure[]{\includegraphics[width=.44\columnwidth]{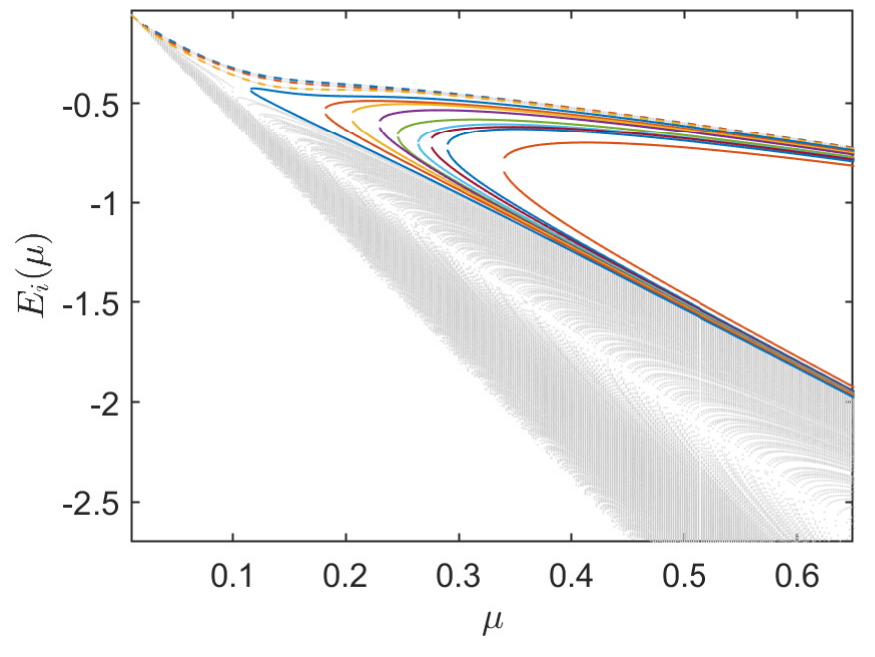}}\\
~	\subfigure[]{\includegraphics[width=.44\columnwidth]{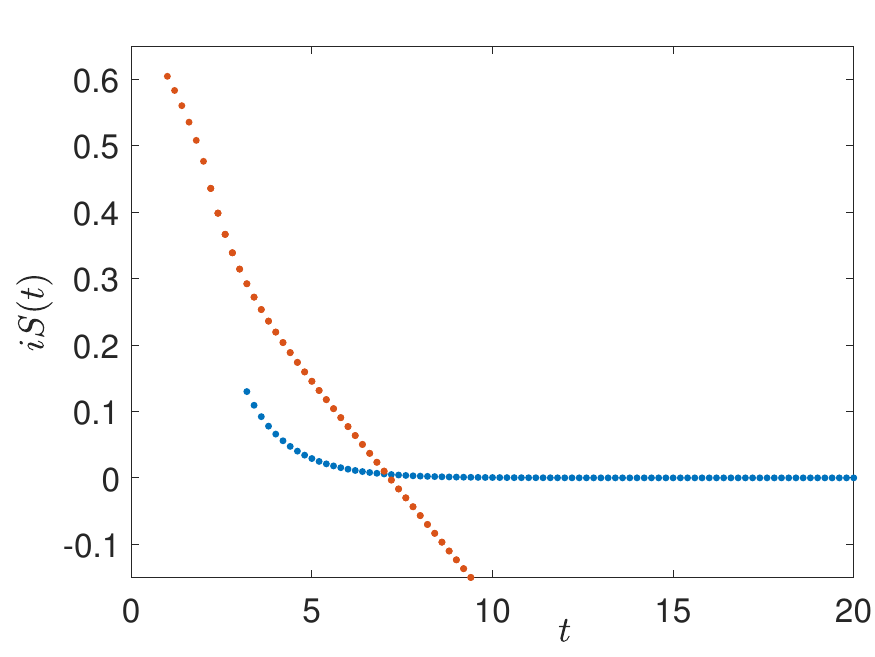}}
~~~	\subfigure[]{\includegraphics[width=.44\columnwidth]{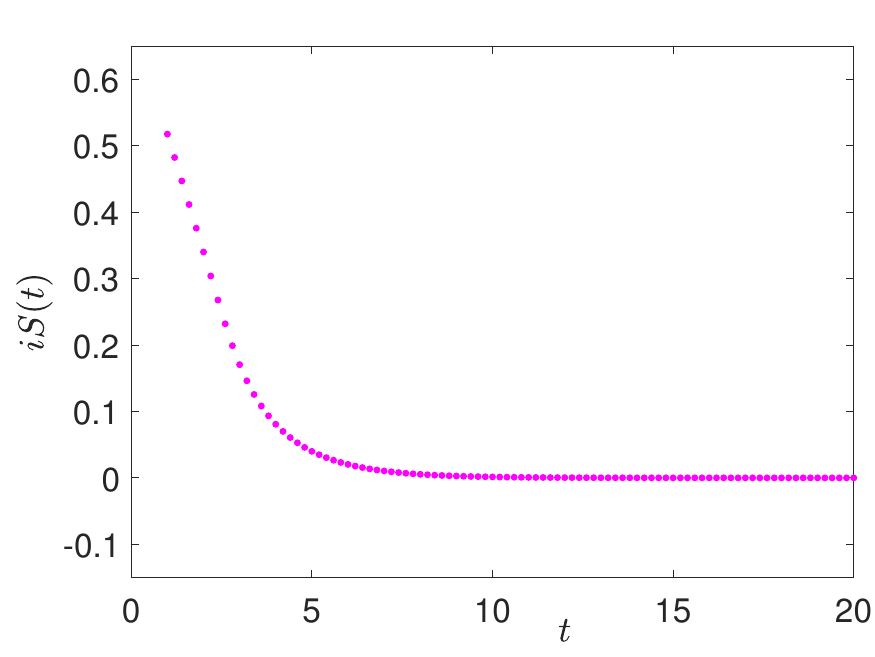}}
	%	\subfigure[]{\includegraphics[width=.45\columnwidth]{NEPs.eps}}
	\caption{  %It is seen that the cross of wormhole and black hole branch indicate a first order transition. We also observe a second order phase transition around $\mu=0.3$, where these two branches merge. Moreover, the merging point in the figure is also shown as the converging point of lines of three different color. For large $\mu$, this second order phase transition will become a crossover completely. 
		(a) Real part of the eigenvalues of the Liouvillian Eq.~(\ref{eq:lio}) for a disorder realization with $N=12$ in the parity sector that does not include the ground state. The red dots indicate eigenvalues with non-zero imaginary part. The EP's occur when pairs of complex eigenvalues hit the real axis. 	
		(b) Purely real eigenvalues $E_i$ of the Liouvillian Eq.~(\ref{eq:lio}) as a function of $\mu$ for a different disorder realization with $N = 12$ also in the parity sector that does not include the ground state. The colored dashed lines at the top are the intruder purely real eigenvalues, not related to the EP's, which are closest to the steady state. Despite the intruder eigenvalues, the gap is mostly controlled by the repulsion of real eigenvalues coming from the bifurcation of EPs after they hit the real axis.
		%that control the intrinsic relaxation of the system.
		(c)-(d) $iS$ Eq.~(\ref{eq:iS}) in the large $N$ limit, see End Matter for additional details, 
		%, see Ref.~\cite{kawabata2023} for a definition, 
		versus $t$ for $\mu = 0.15$ (c), $\mu = 0.35$ (d). Two saddle-points are identified for $\mu \lesssim 0.3$. For $\mu = 0.15$, the red line describes a short-time phase dominated by the $H_{\rm SYK}$ Eq.~(\ref{eq:Hamiltonian} dynamics while the blue line stands for a long-time phase dominated by the bath leading to the steady state. For $\mu \gtrsim 0.3$, see Fig.~\ref{fig:iStEP}(d), the first order transition terminates, only one saddle point is identified (pink line) and the dynamics is bath dominated at all times. This termination occurs precisely in the range of $\mu \sim 0.3$ at which, according to plots (a), (b) above, a proliferation of EP occurs and the full spectrum close to the steady state becomes purely real.}
	\label{fig:iStEP}
	%, see Fig.~\ref{fig:Minmodel1}. 
	%\amg{All what I write about this in the caption and the main text is wishful thinking but I do not have better numerical arguments.} }
	\end{figure}

We observe qualitative agreement between the spectral gap (right panel) obtained from the spectrum of the Liouvillian and the decay rate of retarded Green's function (left panel). In both cases, the decay rate $\Gamma_0$ has a finite value in the $\mu \to 0$ limit, which for the spectral gap requires the extrapolation of the results obtained at finite $\mu$.  For $\mu \ll 1$, $\Gamma_0$ grows slowly. This the region of system dominated dynamics where thermalization is governed by the system dynamics and not by the bath.
This slow increases ends at a local maximum followed by an abrupt decrease precisely at a value of $\mu \sim 0.1$ at which EP's start to become important, see Fig.~\ref{fig:iStEP}(a)(b) in the region close to zero eigenvalue of the Liouvillian corresponding to the steady state. For larger $\mu$, $\Gamma_0$ increases monotonically at a rate that eventually becomes linear signaling a phase of bath dominated dynamics. %Physically, the local maximum before the decrease for small $\mu \sim 0.07$ represents the region at which EPs begin to alter the lowest eigenvalues of the Liouvillian which induce a 
We note that the EP's decreases the decay rate because once the EP hits the real axis, each eigenvalues moves in opposite directions. The one moving towards smaller values induces a smaller value of the gap.  
 There are also quantitative differences: the position of the minimum in $\Gamma_0(\mu)$ is about a factor $2$ smaller. We do not have a clear understanding of the reasons for this discrepancy. It could be that the system size is still too small for quantitative comparisons with the large $N$ limit, or that the Green's functions does not receive contributions from all the lowest real eigenvalues while the gap is obviously affected by them, see dashed lines in Fig.~\ref{fig:iStEP}(b).  
%Supporting the interpretation of the local maximum of the decay as an exceptional point, a %transition between oscillatory decay and exponential damping also appears in the Keldysh %Green function (see Fig. 2 of Ref. \cite{GarciaGarcia2023}).   

We stress that except for the finite decay rate in the $\mu \to 0$ limit, which is expected because results can only be trusted for $\mu \gtrsim 1/N$ \cite{Mori2024}, the rest of features are present in the simple analytical expression for $N=4$  Eq.~(\ref{eq:gamma0}).  
%In this case, the failing of analytical result to reproduce the expected finite value of $\Gamma_0$ is well understood because, as mentioned earlier,  the .

%The results clearly indicate that the dissipative gap in the thermodynamic limit is, for sufficiently large $\mu$, controlled by an EP whose position induces a non-monotonic change in decay rate as a function to the coupling strength to the bath in a similar way at what we found analytically in the $N=4$ case. In other words, 
In summary, the presence of EP signals   the growing influence of the Markovian bath in the dynamics as the coupling $\mu$ increases.   
%The details, however, vary between the Keldish Green function formalism and the numerical results due to the fact that %results from the Keldish formalism include probably more contributions to the equilibration. 

%It must be noted, however, that once we do the size scaling to the thermodynamic limit, if we extrapolate the linear behavior for weak coupling to $\mu=0$ we obtain a non-zero value $\Gamma(0) \sim 0.45$. As mentioned before this is due to the non-commutativity of the thermodynamic and weak coupling limit and has been also found recently in different models \cite{Mori2024}. The decay rate of the large $N$ Green's function also captures this behavior. 

Although the size scaling of the gap in the thermodynamic limit shows a clear anomalous behavior of the relaxation, the results for finite $N$ are not so neat. 
%We study them in more detail in the supplementary material.
We show the eigenvalues of the Liouvillian in the parity sector that do not contain the steady state for a disorder realization with $N=12$ in Figs.~\ref{fig:iStEP}(a) and the purely real eigenvalues in Fig.~\ref{fig:iStEP}(b) for a different disorder realization as a function of $\mu$. The red dots in the former correspond to eigenvalues with non-zero imaginary parts while the black ones correspond to purely real eigenvalues. We can clearly see the transition from the weak coupling limit to the strong coupling where the eigenvalues are clustered around real values, separated by $\mu$ ($2\mu$ in the figure as it is only a single parity block) \cite{kulkarni2022}, corresponding to the eigenvalues of the bath part of the Liouvillian, $i\mu\sum_i \psi_i^+ \psi_i^-$, which is Hermitian. The bifurcations due to the EPs are clearly seen. After the bifurcation, one of the eigenvalues goes to one of the clusters and the other to the consecutive one. The EPs are therefore marking the transition between a chaotic spectrum corresponding to the weak coupling regime to a regular, nearly equispaced, spectrum corresponding to the strong coupling regime.

In the region close to the zero eigenvalue, relevant to the dissipative gap, EP's start to be relevant around $\mu \sim 0.1$, see Fig.~\ref{fig:iStEP}(b). However, we also observe "intruder" states whose eigenvalues have no imaginary part and are related to the exact eigenstates of the SYK Hamiltonian and therefore capture intrinsic relaxation properties of the system. Although, these states do not show an EP as a function of $\mu$, its behavior is also influenced by the presence of the EP because at the bifurcation, the EP states repel the rest of the eigenvalues and induce the observed non-monotonicity of the gap. 
%A more detailed account of this behavior can be found at the supplementary material \cite{Supplemental}. 
More specifically, we have observed in our numerical calculations that the number of these intruder eigenvalues scales with $d$, the dimension of the Hilbert space of the closed SYK model, while the total number of eigenstates of the Liouvillian scale with $d^2$.   
As a consequence, they lose importance in the thermodynamic limit and we recover the clear anomalous relaxation observed in Fig.~\ref{fig:gap_tl} for both the decay of Green's function and the dissipative gap.

%\begin{figure}[htb]
%	\centering
%	\includegraphics[width=0.99\columnwidth]{re4.pdf}
%        \includegraphics[width=0.99   \columnwidth]{gapfig.pdf}
%	\caption{Real part of the eigenvalues for a numerical example with $N=12$. The red lines indicate eigenvalues with non-zero imaginary part.} 
%	\label{fig:N12full}
%\end{figure}  

So far, we have focused on the EP closer to the steady state related to the longest time scales. We now explore the role of higher energy EP's related to shorter time scales. For that purpose, we compute $iS(t)$ \cite{kawabata2023} for the Liouvillian Eq.~(\ref{eq:lio}),
%see Ref.~\cite{kawabata2023} for a definition, 
\begin{equation}
iS(t)=\lim_{N\rightarrow \infty} \frac{\ln {\mathrm Tr}_{\cal{H}^+ \otimes \cal{H}^-} e^{t\cal{L}}}{N},
\label{eq:iS}
\end{equation}
which is an analogue of the Loschmidt echo for dissipative
quantum systems as it describes the overlap between initial and time evolved states of the vectorized density matrix. In the End Matter, we provide details for its calculation in the large $N$ limit. We note that a very similar quantity, termed free energy, with the same phase diagram was computed earlier in  Ref.~\cite{GarciaGarcia2023}. % \ref{sec:endmatter}
% after the substitution $\beta = t$ with $\beta$ the inverse temperature} \textcolor{red}{its evaluation becomes feasible in the large N limit, attributed to its analogue to partition function of a non-Hermitian model \cite{GarciaGarcia2023}. More calculational details are given in appendix.}. 
$iS(t)$ in this model shows \cite{kawabata2023} a rich pattern of dynamical phase transitions depending on  $\mu$. A first order transition is observed for $\mu \lesssim 0.3$ that becomes a crossover at a certain $\mu_c \sim 0.3$, see  Fig.~\ref{fig:iStEP}(a) and (b). The flat $iS$ for small $\mu$ and long times is related to the existence of the so called Keldysh wormholes solution in the large $N$ analysis of Ref.~\cite{GarciaGarcia2023} while the short-time dynamics is controlled by the black hole phase. 
%The latter describes low energy properties of the system and therefore, in our context, if dominant, it indicates that the out of equilibrium dynamics is still mostly governed by the system and not by the bath despite the existence of EP close to the steady state. 
The observed change of behavior of $iS(t)$ as $\mu$ is increased is associated with the already commented change in the spectrum, from chaotic to regular, and from complex to real, as a function of $\mu$. This change is governed by the EPs. Indeed, the value of $\mu_c$ also coincides roughly with the proliferation of EPs seen in Fig.~\ref{fig:iStEP}(a) and (b) and specifically the absence of complex eigenvalues in the spectrum close to the steady state. 

A natural question to ask is about the universality of our results. 
%We note that the relevance of EP's in the equilibration process probably require many-body quantum chaos and likely strong interactions. %The termination of the first order transition in the return probability at a certain $\mu$ signaling a bath driven time evolution is bound from below by the value of $\mu$ at which the decay rate has a minimum. 
%This calls the question of the degree of universality of our results. Does the non-monotonicity of the decay rate with the coupling depends on the choice of jump operators?, is it specific to the SYK model? 
We expect some dependence on the bath because it could effectively add a non-trivial non-Hermitian part to the Hamiltonian  which is a constant in our case. 
Having said that, we think that for sufficiently strong interactions, leading in general to quantum chaotic dynamics, we would still observe a weak coupling region dominated by the system dynamics that transits towards a bath dominated phase characterized by the EPs.  
%The coupling $\mu$ could be made random as 
%The same applies to the choice of the model, the SYK in this case.  
%The dependence of the results on $q$ suggests that while some monotonicity will remain in most cases, 
%Although, the details of the dependence of the decay rate on $\mu$ are likely model dependent but, provided the dynamics is quantum chaotic, we still expect qualitatively similar features such as the non-monotonicity of the decay rate as the bath coupling increases. 

%Finally, we note the possible existence of a Keldysh wormhole gravity dual of our SYK setting, likely related to a near deSitter background in two dimensions \cite{turiaci2019,cotler2020} opens the intriguing possibility that EP may play a role in holography and quantum gravity as well. 

In conclusion, we have shown that EP's have a profound impact in the out of equilibrium dynamics of the SYK model coupled to a Markovian environment. The bifurcation behavior related to the exceptional points governs the transition between the weak coupling region dominated by the intrinsic relaxation of the many-body chaotic system, and the strong coupling regime dominated by the bath dynamics inducing both an anomalous relaxation where the dissipative gap is reduced as the coupling is increased and, for a slightly stronger coupling, the termination of dynamical phase transitions separating a short-time bath driven phase from a late-time system dominated phase. 
%Our results paves the way for a more detailed understanding of the dynamics of dissipative many-body quantum chaotic systems by the study of EP's which are much easier to study and compute than observables involving the calculation of eigenvectors. 

\begin{acknowledgments}
AMGG thanks illuminating discussions with Lucas Sa and Jac Verbaarschot, especially regarding the definition of the spectral gap. AMGG and JPZ were partially supported by the National Science Foundation of China (NSFC): Individual Grant No. 12374138, Research Fund for International Senior Scientists No. 12350710180. JD and RAM acknowledge financial support by Projects No. PID2022-136285NB-C31 and PID2022-136992NB-I100
funded
by MCIN/AEI/10.13039/501100011033 and FEDER ”A way of making Europe”
\end{acknowledgments}
\bibliography{libryu}

\appendix 

\section*{End Matter}\label{sec:endmatter}

\section{Details of dissipation form factor and relaxation rate }\label{app:dff}
This appendix aims to provide details for the calculation of both $iS$ Eq.~(\ref{eq:iS}) and the Green's function decay $\Gamma_0$ in the large $N$ limit presented in the main text.  

We recall that $iS(t)=\lim_{N\rightarrow \infty} \frac{\ln \mathrm{Tr}_{\cal{H}^+ \otimes \cal{H}^-} e^{t\cal{L}}}{N}$, where $\mathrm{Tr}_{\cal{H}^+ \otimes \cal{H}^-} e^{t\cal{L}} = \sum\langle j,i|e^{t\cal{L}}|i,j\rangle $ is known as the dissipative form factor (DFF) \cite{kawabata2023}. In this expression, $|i,j\rangle \in \cal{H}^+ \otimes \cal{H}^-$ are a basis of the vectorized density matrix $|\rho\rangle$, obtained from the Choi-Jamiolkowski isomorphism $\rho=\sum \rho_{ij}|i\rangle \langle j| ~\to~ |\rho\rangle=\sum \rho_{ij}|i,j\rangle$. Meanwhile, this vectorization procedure maps the Liouvillian $\cal{L}[\rho]$ in Eq.~(\ref{eq:SYKLindblad}) to the operator $\cal{L}$ in Eq.~(\ref{eq:lio}). 
% DFF measures the change of a state $|\rho\rangle$ during an evolution, such that it monitors the dissipation process for the open system\cite{GarciaGarcia2023,kawabata2023}. 
Its late-time dynamics approaches a steady state \cite{kulkarni2022,GarciaGarcia2023} corresponding to a thermo-field double state $|\rho\rangle =\sum_i |i,i\rangle$ at infinite temperature.

Note that the DFF is analogue to a partition function ${\mathrm Tr}[e^{-t H}]$ with $H=-\cal{L}$ a non-unitary Hamiltonian, and $t$ the inverse temperature. This similarity allows us to employ techniques developed for computing thermodynamic properties of Hermitian SYK models \cite{maldacena2016,maldacena2018} in the study of this dissipative quantum dissipation problem. 

In the large-$N$ limit, a possible way to compute the DFF is by writing it as a path integral $\int\! D\psi e^{iNS}$, with $S$ the action, which is evaluated by the saddle-point method. Recalling $\cal{L}$ in Eq.~(\ref{eq:lio}), $S$ is  in our case given by,    
\begin{widetext}
\begin{align}
iS =& \int_{0}^{t} d\tau [ -\frac{1}{2}\sum_{i} (\psi^{+}_{i}\partial\psi^{+}_{i} + \psi^{-}_{i}\partial\psi^{-}_{i} )  - i\mu \sum_i \psi^+_{i} \psi^-_{i} -\mu N   \notag\\ 
&\qquad\qquad\qquad  -i^{q/2+1}\sum J_{i_1\cdots i_q} \psi^+_{i_1} \cdots \psi^+_{i_q}  +  i^{q/2+1}\sum i^q J_{i_1\cdots i_q} \psi^-_{i_1} \cdots \psi^-_{i_q}  ].
\end{align}
\end{widetext}
Superscripts ``$\pm$'' in $\psi_i^{\pm}$ labels two kinds of particles obtained from the vectorization procedure. The vectorization method reduces the direct time evolution of the density matrix to a simpler path integral problem, but at the cost of doubling the degrees of freedom.

The standard method for computing $iS$ includes several steps: first we carry out an annealed average of $\int D\psi e^{iNS}$ over random couplings $J_{i_1\cdots i_q}$, then we integrate out fields $\psi_i$, after introducing two bilocal fields $G_{ab}$ and $\Sigma_{ab}$, which results in the following action,
%\begin{widetext}
\begin{align}
iS=& \frac{1}{2}\log\det(\delta_{ab}\partial-\Sigma_{ab}) - \frac{i \mu}{2}\int_0^t d\tau( G_{+-}(\tau,\tau)-G_{-+}(\tau,\tau))  \notag\\
& -\frac{1}{2}\sum_{ab}\int_0^t \!\!\!\int_0^t d\tau     d\tau'[\Sigma_{ab}(\tau,\tau')G_{ab}(\tau,\tau') +\frac{J^2}{q} t_{ab}  G_{ab}(\tau,\tau')^q ],  
\end{align}
%\end{widetext}
where $t_{++}=t_{--}=1$, $t_{+-}=t_{-+}=-(-1)^{q/2}$, and $a,b\in\{+,-\}$. 

In the large-N limit, the on-shell $G_{ab}$ satisfies the definition
\begin{align}
G_{ab}(\tau) = \frac{1}{N}\sum_i \langle \psi_i^a(\tau) \psi_i^b(0) \rangle, 
\end{align}
where $G_{ab}$ and $\Sigma_{ab}$ are the Green's function and self-energy, respectively. Besides, applying the variation to $iS$ with respect to $\delta G_{ab}$ and $\delta \Sigma_{ab}$ gives rise to the saddle-point Schwinger-Dyson equations,
\begin{align}
& \partial G_{++}-\Sigma_{++}*G_{++}-\Sigma_{+-}*G_{-+}=\delta(\tau), \notag\\
& \partial G_{+-}-\Sigma_{++}*G_{+-}-\Sigma_{+-}*G_{--}=0, \notag\\
& \Sigma_{++} = -J^2 G_{++}^{q-1} ,\qquad \Sigma_{+-}=(-1)^{q/2}J^2 G_{+-}^{q-1} - i \mu \delta(\tau), 
\end{align}
in which ``$*$'' denotes convolution. $G_{ab}$ is then obtained by solving these equations numerically. 
In the last step, these solutions are plugged into the action above which leads directly to the desired $iS(t)$ depicted in Fig.~\ref{eq:iS}.

 %then we can evaluate $\int D\psi e^{iNS}$ in the large-N limit through employing saddle-point approximation, by inserting the on-shell $G_{ab}$ back into the action $iS$. 

%Note that $G_{ab}$ is not exactly the real-time Green's function defined in Keldysh formalism, which are generally used in the study of the out-of-equilibrium dynamics. But at late times, when the system reaches to the steady state, their relation are built, and 

Regarding the calculation of $\Gamma_0$ in Fig.~\ref{fig:gap_tl} (left), we note that, for sufficiently long times, close to the steady state,  $G_{+-}(\tau) = G^<(\tau)$ and $G_{-+}(\tau) = G^>(\tau)$. In this limit, the exponential decay rate $\Gamma_0$ of $G_{++}(\tau) \approx -i G_{+-}(\tau)$ $\in \mathbb{R}$ ($\tau\in [0,t/2]$) reflects the relaxation properties for the open SYK system. We have observed \cite{GarciaGarcia2023} that $G_{ab}(\tau)$ exhibit an exponential decay with oscillations superimposed for $\mu < 0.15$. In order to compute $\Gamma_0$ as a function of $\mu$ in Fig.~\ref{fig:gap_tl}, we have fitted the numerical results to  
\begin{equation}\left\{\begin{aligned}
& G_{++}(\tau) = A e^{-\Gamma_0 \tau} \sin (\Omega \tau +b), \qquad \mu<0.15  \notag\\
& \ln|G_{++}(\tau)| = -\Gamma_0 \tau +c, \qquad \mu\geq 0.15.
\end{aligned}\right.\end{equation}

\end{document}